\documentclass{iopart}
\usepackage[utf8]{inputenc}
\usepackage[T1]{fontenc}
\usepackage[english]{babel}
\usepackage[colorlinks,citecolor=blue]{hyperref}
\usepackage{times}
\usepackage[sort&compress,numbers]{natbib}
\usepackage{graphicx}

\providecommand{\ndensity}[1]{\ensuremath{n_{#1}}}
\providecommand{\ldensity}[1]{\ensuremath{l_{#1}}}
\providecommand{\tdensity}[1]{\ensuremath{l_{#1}}}

\providecommand{\nc}{\ndensity{C}}
\providecommand{\nd}{\ndensity{D}}
\providecommand{\cc}{\ldensity{CC}}
\providecommand{\dd}{\ldensity{DD}}
\providecommand{\cd}{\ldensity{CD}}
\providecommand{\avk}{\ensuremath{\langle k \rangle}}

\begin{document}
\title[A homoclinic route to asymptotic full cooperation]{A homoclinic route to
asymptotic full cooperation in adaptive networks and its failure}
\author{Gerd Zschaler$^1$, Arne Traulsen$^2$ and Thilo Gross$^1$}
  \address{$^1$Max-Planck-Institut für Physik komplexer Systeme,\\ Nöthnitzer
Str. 38, 01187 Dresden, Germany}
  \address{$^2$Max-Planck-Institut für Evolutionsbiologie,\\
August-Thienemann-Str. 2, 24306 Plön, Germany}
\ead{zschaler@pks.mpg.de}

\begin{abstract}
We consider the evolutionary dynamics of a cooperative game on an adaptive
network, where the strategies of agents (cooperation or defection) feed back on
their local interaction topology. While mutual cooperation is the social
optimum, unilateral defection yields a higher payoff and undermines the
evolution of cooperation. Although no \emph{a priori} advantage is given to
cooperators, an intrinsic dynamical mechanism can lead asymptotically to a state
of full cooperation. In finite systems, this state is characterized by long
periods of strong cooperation interrupted by sudden episodes of predominant
defection, suggesting a possible mechanism for the systemic failure of
cooperation in real-world systems.
\end{abstract}

\submitto{\NJP}

\maketitle

\section{Introduction}
Understanding the evolution of cooperation between selfish players is of
importance not only because of its central role in biology and the social
sciences \cite{Nowak2004, Macy2002} but also to address the emergence and
failure of cooperation in human societies \cite{Turchin2003}. Previous work has
identified
several key mechanisms \cite{Nowak2006} leading to the evolution and fixation of
cooperation. Among others, it was found that spatial structure can promote
cooperation \cite{Axelrod1984} when agents are placed on regular lattices
\cite{Nowak1992} or on complex networks \cite{Szabo2007, Ohtsuki2006,
Santos2005}. Perhaps the newest development in this direction is the
investigation of games on adaptive networks \cite{Gross2008}, in which the
players' behaviour feeds back on the network topology. Cooperation is promoted
if cooperating players can secure an advantageous topological position, directly
due to avoidance of defectors \cite{Santos2006, Tanimoto2007, Hanaki2007,
Fu2008, Fu2009, Chen2009, Szolnoki2009}, or indirectly due to the continuous
arrival of new players \cite{Poncela2009} or other ongoing changes of the
topology \cite{Szolnoki2008, Szolnoki2009a}. Furthermore, cooperation in
adaptive networks can profit from the emergence of a self-organized leadership
structure \cite{Zimmermann2000, Zimmermann2004, Zimmermann2005, Eguiluz2005} and
the formation of strongly heterogeneous topologies \cite{Ebel2002, Ren2006,
Poncela2008}. Finally, it was shown that cooperation on adaptive networks can
result in an arms race for swift topological response that promotes cooperation
\cite{Segbroeck2008, Segbroeck2009}. Interestingly, full cooperation of every
single agent has been observed in a number of studies. This is explained in part
in \cite{Pacheco2006a}, showing that in the limit of fast rewiring, cooperative
games on adaptive networks can be mapped on coordination games in a well-mixed
population.
The coevolutionary games listed here and other forms thereof have been reviewed
in \cite{Perc2010}.

Although cooperation on adaptive networks is an inherently dynamical process,
previous studies in this field have focused mainly on the average, and therefore
quasi-static, level of cooperation that evolves. Dynamical phenomena such as the
appearance of oscillations are known to play a role in non-adaptive models of
iterated and cyclic games, where they have been investigated thoroughly (e.g.,
\cite{Nowak1989, Imhof2005, Reichenbach2006}). Furthermore, interesting
dynamical behaviour has been observed already in a pioneering publication on
adaptive networks \cite{Zimmermann2000} and many subsequent works
\cite{Zimmermann2000, Zimmermann2004, Zimmermann2005, Eguiluz2005, Hanaki2007,
Szolnoki2008, Szolnoki2009} and has been analysed to some extent
in \cite{Suzuki2008}.
Only recently, oscillations have been observed in an adaptive network model,
where they were interpreted in terms of a Red Queen mechanism that is also able
to promote cooperation \cite{Szolnoki2009a}.

In the present paper, we study an evolutionary game on an adaptive network where
the collective dynamics of the agents lead to asymptotic full cooperation. In
contrast to previous work that mainly focused on local update rules based on the
payoff of neighbouring nodes \cite{Zimmermann2000, Zimmermann2004,
Zimmermann2005, Santos2006}, the agents use non-local information about the
general performance of the strategies in our model. This corresponds to the
accessibility of certain global knowledge in social systems through, e.g.,
the media.

Investigating the dynamics of the present model, we find oscillations in the
number of cooperating players, in which the time-averaged payoff of the latter
equals that of the defectors in the limit of infinite population size.
Nevertheless, a state of full cooperation is approached asymptotically if the
rate of topological change exceeds a finite threshold. This is made possible by
a dynamical mechanism involving the formation of a homoclinic loop in a global
bifurcation. Combining ideas from network science, nonlinear dynamics, and
evolutionary game theory, we derive a low-dimensional analytical model
explaining this dynamic phenomenon far from equilibrium. Furthermore we show
that in finite populations, this mechanisms can lead to periods of almost full
cooperation interrupted by recurrent collapses to episodes of predominant
defection, revealing a scenario for the sudden failure of cooperation.

\section{Model}
We consider an undirected network of $N$ nodes, representing agents, and $K$
links, representing interactions. Each agent $i$ is assigned a strategy
$\sigma_i$, which can either be cooperation, $C:=1$, or defection, $D:=2$. The
payoff gained in an interaction is modelled by the snowdrift game, a
paradigmatic model in game theory \cite{Doebeli2005}. In a common
parametrization, two interacting agents receive a benefit $b$ if either of them
cooperates. Cooperation incurs a cost $c<b$, which is divided between
cooperators, but not defectors. The payoff a player $i$ receives from the
interaction with player $j$ can then be written as $M_{\sigma_i\sigma_j}$, where
\begin{equation}
\label{eq:po-matrix}
\mathrm{\mathbf{M}} = \left(\begin{array}{cc} b - \frac{c}{2} & b-c \\ b &
0\end{array} \right)
\end{equation}
is the payoff matrix. The total payoff player $i$ gains from all interactions is
given by $\pi_i=\sum_{j\sim i} M_{\sigma_i\sigma_j}$, where the summation runs
over all $j$ linked to $i$.

Starting from a random graph and randomly assigned equiprobable strategies, we
evolve the network as follows: In every time step, one link is selected at
random. With probability $p$, this focal link is rewired. Otherwise, i.e., with
probability $q=1-p$, one of the linked players adopts the other player's
strategy. For large $p$, players thus tend to change their interaction partners,
whereas for small $p$ they tend to revise their behaviour.

To complete the model, we have to specify which agent copies the other's
strategy in a strategy adoption event and which agent keeps the link when
rewiring takes place. Most previous models assume that the agents' access to
information is governed by the same network as the underlying games, forcing the
agents to base their decisions on information from direct neighbours.
The same network topology thus determines three different aspects of the system:
the interaction partners of an agent, against whom the game is played, the
potential role models, whose strategies can be adopted, and the agents from
which information can be obtained. However, for intelligent agents, and
especially humans, there is no reason to assume that these three network roles
are all fulfilled by coinciding topologies \cite{Ohtsuki2007a}. In fact, the use
of identical networks for information and state transmission has recently been
criticized for the related application of epidemic spreading \cite{Funk2009,
Funk2010}.

In the present paper, we assume that information transfer in the population is
not governed exclusively by the interaction network. As a first approximation,
we consider the simplest case in which the information transmission network is
replaced by an effective global coupling, as information can be rapidly
transmitted and in the human population is also transported by the mass media.
In the main part of this work, we assume that the agents rely on their
perception of a strategy's \emph{general} performance. This global measure for a
strategy $\sigma$ is obtained as the average payoff of all agents currently
using $\sigma$,
\begin{equation}
\label{eq:fitness}
\phi(\sigma) = \sum_{i, \sigma_i = \sigma} \frac{\pi_i}{\ndensity{\sigma}N} =
\sum_{\sigma' \in
\left\{C, D\right\}} (1+\delta_{\sigma\sigma'}) M_{\sigma\sigma'}
\frac{\ldensity{\sigma\sigma'}}{\ndensity{\sigma}},
\end{equation}
where $\delta$ is the Kronecker delta, $\ndensity{\sigma}$ is the fraction of
the population using the strategy, and $\ldensity{\sigma\sigma'}$ is the number
of links between agents using strategies $\sigma$ and $\sigma'$ normalized by
$N$. The validity of assuming such a measure based on global information is
discussed in section \ref{sec:analytics}, where we consider the effect of a
finite ``information horizon'' within which the agents assess the performance of
a strategy. If strategy adoption occurs on a link connecting the agents $i$ and
$j$, then we assign the strategy of agent $i$ to agent $j$ with probability
given by the Fermi function \cite{Blume1993, Szabo1998}
\begin{equation}
 \label{eq:fermi}
  f_{\beta}(i,j) = \left(1+e^{-\beta\left[\phi(\sigma_i) -
\phi(\sigma_j)\right]}\right)^{-1}.
\end{equation}
Otherwise, i.e., with probability $f_{\beta}(j,i)=1-f_{\beta}(i,j)$, the
strategy of agent $j$ is assigned to agent $i$. The parameter $\beta$ is the
selection intensity and corresponds to an inverse temperature. For small $\beta$
the strategy adoption is almost random, whereas for $\beta\to\infty$ the more
successful strategy is always adopted.
Following established practice \cite{Nowak2004a, Ohtsuki2006,
Ohtsuki2007a}, we mainly focus on the case of weak selection (small $\beta$),
which is known to be highly relevant for biology and supported by recent
evidence for social systems \cite{Traulsen2010}. Weak selection also implies
that an agent's local neighbourhood is more important for an agent's strategy
than the global information. While the global information (weakly) influences
the probability that the agent's strategy changes in a given strategy update,
the local neighbourhood governs the rates at which update events involving the
focal agent occur.

In a similar fashion, we assume that the players using the more successful
strategy are more likely to keep links during rewiring events. If a rewiring
event occurs on a link connecting the agents $i$ and $j$, the link is cut and
then a new link is established between a randomly selected agent $k$ and agent
$i$ (with probability $f_{\alpha}(i,j)$) or between $k$ and $j$ (with
probability $f_{\alpha}(j,i)$). Here, we have used the Fermi function with
selection intensity $\alpha$ to capture that agents following a successful
strategy may find it easier to attract new contacts. Using the average payoff
$\phi(\sigma)$ instead of the payoff obtained by individual agents prevents
successful players from acquiring an unrealistically large number of links,
which could otherwise lead to the formation of star-like topologies.
It also implies that all agents using the same strategy are considered
equivalent for the dynamics. We note that a similar assumption is made in the
large class of models where the topological processes depend exclusively on the
nodes' states instead of their fitness (e.g., \cite{Pacheco2006a,
Segbroeck2009}).

\section{Simulation results}
\begin{figure}
\centering
\includegraphics[width=0.65\textwidth]{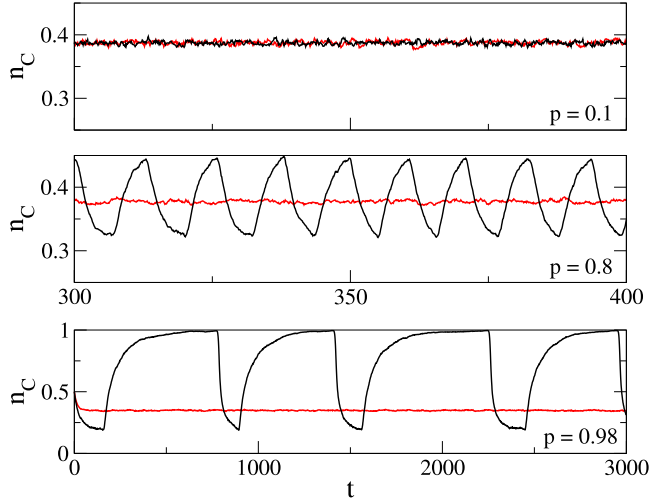}
\caption{\label{fig:sim-oscill}
Time series of the fraction of cooperators in an adaptive network for different
rewiring rates $p$. When rewiring occurs almost at random ($\alpha=0.1$, red),
cooperation and defection coexist at a stationary level for all $p$. When
players following the more successful strategy keep the link with high
probability ($\alpha=30$, black), oscillations appear as the rate of rewiring
exceeds a critical threshold. Parameters: $N=10^5$, $K=10^6$, $\beta=0.1$,
$b=1$, $c=0.8$.}
\end{figure}

In order to explore the dynamics of the model, we run individual-based
stochastic simulations for $N=10^5$ and $K=10^6$. Typical time series for
different rewiring rates $p$ are shown in \fref{fig:sim-oscill}. For weak
selection ($\alpha, \beta\ll 1$), the system approaches a stable steady state
where both strategies coexist. In this regime the stationary density of
cooperators depends only weakly on $p$. If rewiring is strongly selective
($\alpha\gg\beta$), then the dynamics depend strongly on $p$. Stationary
behaviour is still observed if $p$ is small, but as $p$ increases, the
system undergoes a continuous transition in which the density of cooperators
starts to oscillate. The same transition can also be observed for higher
$\beta$, but is shifted to greater values of $p$. As $p$ is increased further,
the amplitude and period of the cycle grows. At higher $p$, long periods of
almost full cooperation appear, which are interrupted by sudden episodes of
defection.

\begin{figure}
 \centering
 \includegraphics[width=0.65\textwidth]{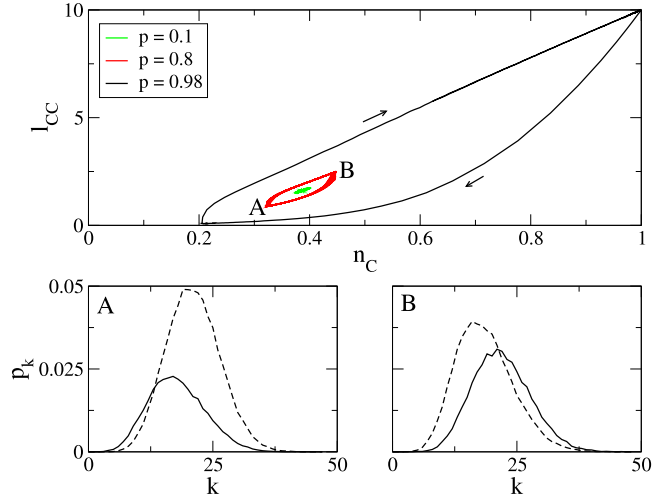}
 \caption{\label{fig:cycle} Cycles of cooperation and defection in agent-based
simulations. Top: dynamics in the $\nc$-$\cc$-plane for rewiring selection
intensity $\alpha=30$. Bottom: degree distributions of cooperators (\full) and
defectors (\dashed) for rewiring rate $p=0.8$ at the two turning points of the
cycle. Simulation parameters as in \fref{fig:sim-oscill}.}
\end{figure}

To understand the onset of oscillations, note first that in the stationary state
the average payoffs of cooperators and defectors have to be identical, so that
strategy adoption and rewiring happen randomly. If, due to fluctuations,
cooperators receive a slightly higher payoff than defectors, then the density of
C-C links, $\cc$, starts to increase due to the effect of the strongly selective
rewiring, which tends to de-mix the network by accumulating links in the
population with the higher payoff. The increasing $\cc$ constitutes a positive
feedback increasing the payoff of the cooperators further. As agents adopt the
cooperating strategy, the system approaches a state where both $\nc$ and $\cc$
are high (point B in \fref{fig:cycle}). In this state, strategy adoption can
overcome the de-mixing effect of rewiring because adoption of the defecting
strategy by cooperators creates many $C$-$D$ links. The payoff of defectors
rises rapidly, leading the system back to a mixed state, where a substantial
number of agents are defecting.

\section{Analytical results}
\label{sec:analytics}
To gain a deeper understanding we formulate a low-dimensional approximation of
our model. The simplest possible model, a mean-field approximation of $\nc$,
yields a one-dimensional ordinary differential equation (ODE), which cannot
reproduce oscillatory long-term dynamics. We therefore use a moment-expansion
approach \cite{Gross2006}, which describes the system on the level of the
density of nodes and links. We treat $\nc$, $\cc$, and $\dd$ as dynamical
variables, whereas $\nd$ and $\cd$ are given by the conservation laws
$\nc+\nd=1$ and $\cc+\dd+\cd=\avk/2$, where $\avk=2K/N$ denotes the mean
degree.

The process that is most difficult to capture in the model is strategy
adoption, because it does not only affect the focal link, but also all other
links connecting to the agent whose strategy is changed. For instance, the
average number of C-C links that are destroyed when a cooperator adopts the
strategy of a defector depends on $\tdensity{DCC}$, the number of D-C-C triplets
per agent. To close the system we use the pair approximation $\tdensity{XYZ} =
(\eta_{XY}\eta_{YZ}/\eta_{XZ}) \ldensity{XY}\ldensity{YZ}/\ndensity{Y}$, where
$\eta_{AB}=1+\delta_{AB}$ denotes factors arising from symmetry. Using this
procedure we obtain
\begin{eqnarray}
\label{eq:coop}
\frac{\rm d}{\rm dt} \nc &= q \cd \left(f_{\beta}- \bar f_{\beta}\right), \\
\label{eq:cclinks}
\frac{\rm d}{\rm dt} \cc &= p \nc \cd f_{\alpha} - p \nd \cc \nonumber \\
& + q \cd \Bigl(1+\frac{\cd}{\nd}\Bigr) f_{\beta} - 2q\frac{\cc \cd}{\nc}\bar
f_{\beta}, \\
\label{eq:ddlinks}
\frac{\rm d}{\rm dt} \dd &= p \nd \cd \bar f_{\alpha} - p \nc \dd \nonumber \\
& + q \cd \Bigl(1+\frac{\cd}{\nc}\Bigr) \bar f_{\beta} - 2q\frac{\dd \cd}{\nd}
f_{\beta},
\end{eqnarray}
where we have introduced the abbreviated notation $f_{\xi} =
\left(1+e^{-{\xi}\left[\phi(C) - \phi(D)\right]}\right)^{-1}$ and $\bar f_{\xi}=
1- f_{\xi}$. In \eref{eq:coop}, the first factor, $q \cd$, denotes the rate of
strategy adoption events while the second factor is the expected change in $\nc$
in each such event. Analogously, the first two terms in \eref{eq:cclinks} and
\eref{eq:ddlinks} describe the gain and loss rates of the respective link
density due to rewiring, while the third and fourth terms account for the link
creation and loss due to strategy adoption.

In the regime of weak rewiring selection, $\alpha \ll 1$, and weak strategy
selection, $\beta \ll 1$, our model reduces to previously studied systems in two
important limiting cases. For fast rewiring ($p\approx 1$), which is almost
random when $\alpha$ is small, the system evolves according to standard
replicator dynamics in a well-mixed populations. On the other hand, when
strategy adoption is much faster than rewiring, $p \ll 1$, the network is almost
static and the modified replicator equation on graphs applies
\cite{Ohtsuki2006a}.

\begin{figure}
\centering
\includegraphics[width=0.65\textwidth]{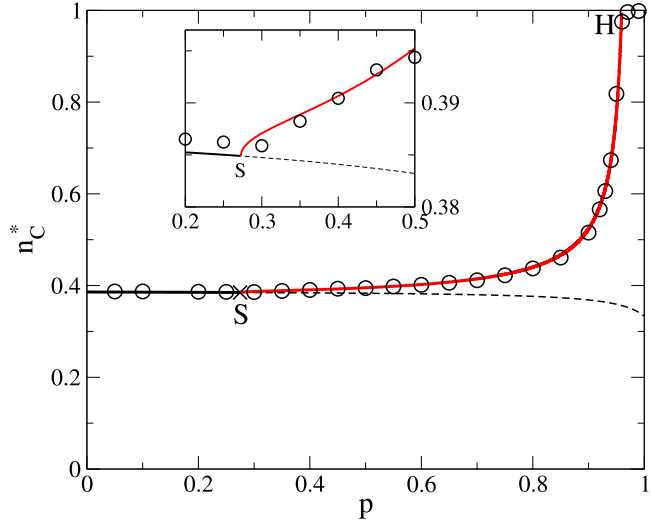}
\caption{\label{fig:bif-p}
Bifurcation diagram for the case of strong rewiring selection ($\alpha=30$). If
rewiring is slow (small $p$), cooperators and defectors coexist in a stable
steady state. At higher rewiring rates the stability is lost in a supercritical
Hopf bifurcation (S), from which a stable limit cycle emerges. The limit cycle
undergoes a homoclinic bifurcation in point H as it connects to a saddle point
at $\nc=1$. The lines show the stable (\full) and unstable (\dashed) steady
state, and the upper turning point of the limit cycle (red), computed in the
low-dimensional model. Circles denote agent-based simulation
results for $N=10^5$, $K=10^6$. The inset shows a blow-up of the bifurcation
point S. See \fref{fig:sim-oscill} for parameters.}
\end{figure}

We now use the low-dimensional model, \eref{eq:coop}--\eref{eq:ddlinks}, to
explore the system with the tools of nonlinear dynamics. A bifurcation diagram
of the ODE system is in good agreement with the results of the agent-based
simulations (\fref{fig:bif-p}). If rewiring is slow (small $p$), the system
approaches an equilibrium in which cooperators and defectors coexist. For
strongly selective rewiring ($\alpha\gg\beta$), however, a critical threshold
$p$ exists at which this steady state is destabilized in a supercritical Hopf
bifurcation (point S in \fref{fig:bif-p}) and a stable limit cycle emerges,
explaining the onset of oscillations.

As $p$ is increased further, the amplitude of the limit cycle grows. Eventually,
the cycle undergoes a homoclinic bifurcation as its upper turning point connects
to the fully cooperative state ($\nc=1$, $\cc=\avk/2$, $\dd=0$). Dynamically,
this state is a saddle point, which the cycle approaches along its stable
manifold and leaves along the unstable manifold, forming a homoclinic loop. In
the saddle point, the velocity at which the system moves along the cycle
approaches zero. If observed at a random point in time, the ODE system is
therefore found to be in the fully cooperative state with probability one.

Let us emphasize that the asymptotic full cooperation is a purely dynamical
effect. The existence of a limit cycle shows that the time-averaged fitness of
cooperators and defectors is equal. However, in the homoclinic bifurcation the
time-average becomes meaningless as it has to be taken over infinite time, while
for any finite time the cooperators dominate.

The deterministic description provided by the ODE system holds in the
thermodynamic limit of large system size. In the agent-based model, full
cooperation is an absorbing state of the strategy dynamics. Small systems can
reach this state and remain at full cooperation. In larger systems ($N>100$),
the system slowly approaches the saddle point along the stable manifold, but is
eventually carried over to the unstable manifold by fluctuations. Once on the
unstable manifold, defection rapidly invades the population, launching the
system into another round on the cycle before the fully cooperative state is
approached again (\fref{fig:cycle}). Because of their stochastic excitable
nature, these invasions of defectors occur at irregular time intervals
(\fref{fig:sim-oscill} bottom), becoming longer with increasing system size.

\begin{figure}
\centering
\includegraphics[width=0.75\textwidth]{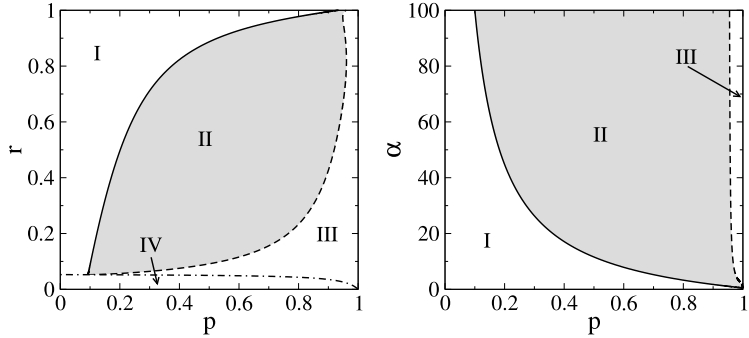}
\caption{\label{fig:bif-2par} Two-parameter bifurcation diagrams showing the
dependence on the rewiring rate $p$ and the cost-to-benefit ratio
\cite{Doebeli2005} $r=c/(2b-c)$ (left, for $\alpha = 30$), and the rewiring
selection strength $\alpha$ (right, for $b=1$, $c=0.8$), resp. In region I,
cooperation and defection coexist in a stationary state. A Hopf bifurcation line
(\full) marks the transition to the oscillatory parameter region (II, shaded),
which is bounded by a line of homoclinic bifurcations (\dashed) leading  to
asymptotic full cooperation (III). Stable full cooperation (IV) is reached via a
transcritical bifurcation (\chain) if $r$ is low. All bifurcation lines
meet in a codimension-2 Takens-Bogdanov bifurcation. See \fref{fig:bif-p}
for additional parameters.}
\end{figure}

In the two-parameter bifurcation diagrams shown in \fref{fig:bif-2par}, the
Hopf and homoclinic bifurcation points form lines, which separate parameter
regions of qualitatively different long-term dynamics. The oscillatory dynamics
and asymptotic full cooperation can be observed in a large parameter region,
demonstrating the robustness of the observed phenomenon to the snowdrift game
conditions (parametrized by the cost-to-benefit ratio $c/(2b-c)$ of mutual
cooperation) and the rewiring selection strength $\alpha$. For larger $\alpha$,
the Hopf bifurcation occurs already at slower rewiring rates
(\fref{fig:bif-2par}, right), whereas the opposite is true for larger $\beta$
(not shown): Although stronger replacement selection leads to oscillations at
higher frequencies than before, it can be balanced by sufficiently fast rewiring
and does not necessarily counteract the homoclinic mechanism.

We note that the homoclinic bifurcation line in \fref{fig:bif-2par} (right)
approaches $p=0.952\approx 20/21$ for large $\alpha$. On this line, 20 out of 21
network updates are rewiring events. Each rewiring event affects only a single
link, while every strategy adoption event affects approximately $\avk = 20$
links. Therefore, at $p=20/21$ the links are in average affected at the rate
$(1-p)\avk=p$ by strategy adoption and rewiring events, i.e., both processes
operate on the same time scale.

\begin{figure}
\centering
\includegraphics[width=0.65\textwidth]{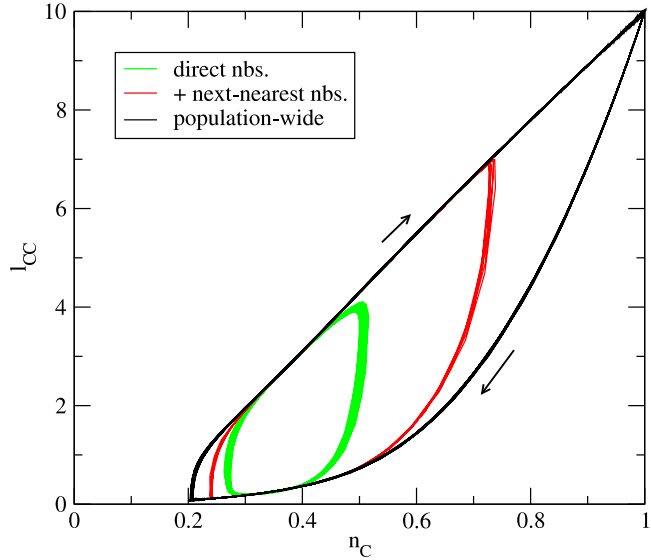}
\caption{\label{fig:neighborhoods} Effect of increasing the agents' information
horizon for fast rewiring in simulations. When the agents assess a strategy's
fitness by averaging over the accumulated payoffs of the focal link's nearest
neighbours only (green), a small limit cycle is obtained. It is larger when also
the next-nearest neighbours are included (red). The size of the cycle grows with
increasing averaging radius. The quasi-homoclinic cycle of long cooperative
periods interrupted by short bursts of defection is recovered when the average
is taken over the whole population (black). Parameters as in
\fref{fig:sim-oscill}, $p=0.98$.}
\end{figure}

Finally, we ask how our findings depend on our initial assumption that the
agents have access to global, i.e., population-wide information in terms of
$\phi(\sigma)$. To that end, we let the players estimate the fitness of
strategies by averaging over a neighbourhood of nodes they can reach in a finite
number of steps rather than over the whole population. Similar behaviour as
before can be observed in simulations with such an ``information horizon''. In
particular, fast selective rewiring leads to oscillations whenever this kind of
neighbourhood information is used. As can be seen in \fref{fig:neighborhoods},
the amplitude and period of these oscillations grow with the information horizon
(for fixed $p$). Therefore, access to the information from a sufficiently large
neighbourhood is, in the present model, necessary to observe the homoclinic
route to full cooperation.

\section{Conclusions}
In this paper, we proposed a model of cooperation on an adaptive network in
which agents have access to non-local information. We observed that full
cooperation is reached asymptotically through a global dynamical
mechanism operating far from equilibrium. In addition to the results shown, we
have verified that similar dynamics can be observed in other parameter regimes,
such as stronger strategy selection, and in variants of the model.

In our model, asymptotic full cooperation is achieved dynamically, rather than
by assembling characteristic topologies that allow cooperators to thrive. In
finite populations, spontaneous collapses of highly cooperative states are
observed. The homoclinic mechanism in combination with noise may thus be a
relevant ingredient in the systemic failure of cooperation that is found in real
world systems.

Although it is conceivable that global information can be accessible in social
systems through, e.g., general beliefs, rumours, or the mass media, we note that
this assumption is not necessary for the oscillatory behaviour. When the agents
are restricted to a finite ``information horizon'', the present model still
exhibits oscillations.

A sufficiently large information horizon is necessary to observe the homoclinic
transition. We note, however, that in a different model in which local update
rules are assumed and which explicitly enforces realistic limits on the node
degrees, a similar transition was found numerically \cite{Szolnoki2009}.
We believe that in the present model, the homoclinic mechanism only requires
large information horizons because they prevent the formation of unrealistic
star-like topologies, which would otherwise stabilize the dynamics.

More work is certainly necessary to explore the role of the homoclinic
mechanism in nature. This work will, however, offer the intriguing possibility
to identify and understand the dynamical features promoting cooperation in many
systems, and may reveal what causes the sudden collapses of cooperative
behaviour observed in the human population.

\ack
We thank C. Huepe and F. Jülicher for helpful comments. A.T. is supported by the
Emmy-Noether program of the DFG.

\bibliographystyle{iopart-num}
\providecommand{\newblock}{}

\end{document}